# If-T: A Benchmark for Type Narrowing

Hanwen Guo[a]  and Ben Greenman[a]

a    University of Utah, Salt Lake City, UT, USA

**Abstract**

**Context**   The design of static type systems that can validate dynamically-typed programs (*gradually*) is an ongoing challenge. A key difficulty is that dynamic code rarely follows datatype-driven design. Programs instead use runtime tests to narrow down the proper usage of incoming data. Type systems for dynamic languages thus need a *type narrowing* mechanism that refines the type environment along individual control paths based on dominating tests, a form of flow-sensitive typing. In order to express refinements, the type system must have some notion of sets and subsets. Since set-theoretic types are computationally and ergonomically complex, the need for type narrowing raises design questions about how to balance precision and performance.

**Inquiry**   To date, the design of type narrowing systems has been driven by intuition, past experience, and examples from users in various language communities. There is no standard that captures desirable and undesirable behaviors. Prior formalizations of narrowing are also significantly more complex than a standard type system, and it is unclear how the extra complexity pays off in terms of concrete examples. This paper addresses the problems through If-T, a language-agnostic *design benchmark* for type narrowing that characterizes the abilities of implementations using simple programs that draw attention to fundamental questions. Unlike a traditional performance-focused benchmark, If-T measures a narrowing system's ability to validate correct code and reject incorrect code. Unlike a test suite, systems are not required to fully conform to If-T. Deviations are acceptable provided they are justified by well-reasoned design considerations, such as compile-time performance.

**Approach**   If-T is guided by the literature on type narrowing, the documentation of gradual languages such as TypeScript, and experiments with typechecker implementations. We have identified a set of core technical dimensions for type narrowing. For each dimension, the benchmark contains a set of topics and (at least) two characterizing programs per topic: one that should typecheck and one that should not typecheck.

**Knowledge**   If-T provides a baseline to measure type narrowing systems. For researchers, it provides criteria to categorize future designs via its collection of positive and negative examples. For language designers, the benchmark demonstrates the payoff of typechecker complexity in terms of concrete examples. Designers can use the examples to decide whether supporting a particular example is worthwhile. Both the benchmark and its implementations are freely available online.

**Grounding**   We have implemented the benchmark for five typecheckers: TypeScript, Flow, Typed Racket, mypy, and Pyright. The results highlight important differences, such as the ability to track logical implications among program variables and typechecking for user-defined narrowing predicates.

**Importance**   Type narrowing is essential for gradual type systems, but the tradeoffs between systems with different complexity have been unclear. If-T clarifies these tradeoffs by illustrating the benefits and limitations of each level of complexity. With If-T as a way to assess implementations in a fair, cross-language manner, future type system designs can strive for a better balance among precision, annotation burden, and performance.

**ACM CCS 2012**
   ▪ **Software and its engineering** → **Control structures; Software notations and tools;** *Functional languages*;

## The Art, Science, and Engineering of Programming

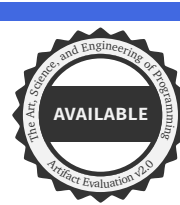 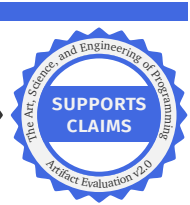

## 1 Introduction

> *"Some account should be taken of the premises in conditional expressions."*
> *John C. Reynolds [57]*

Duck typing is a core competency of dynamic languages. It allows programs to experiment with incoming data, gain partial information, and act on the discovered information without constraining the shape of input data any more than strictly necessary. For example, the Rainfall problem [21, 62] asks for the average rainfall from a list of unreliable weather reports. Any report that does not have a rainfall field, or that has a malformed rainfall value (non-numeric, negative, or greater than 999) should be ignored. The function below, written in a pseudocode inspired by Rhombus [22], Python [74] and TypeScript [73] that we introduce in Section 4.1, cleanly solves the problem using a sequence of conditional tests:

```
1  define avg_rainfall(weather_reports: List(JSON)) -> Number:
2    let total = 0, count = 0
3    for day in weather_reports:
4      if day is Object and has_field(day, "rainfall"):
5        let val = day["rainfall"]
6        if val is Number and 0 ≤ val ≤ 999:
7          total += day["rainfall"]  // expected: no type error, right-hand expression is a number
8          count += 1
9    return (if count > 0: total / count else: 0)
```

Callers of this function are free to send it any sort of JSON data, including data from potentially-flawed sources such as sensors or handwritten spreadsheets. Callers are also free to later modify the shape of the data they send. If, say, a weather-sensor upgrade adds fields or changes the types of unused fields, the function still works the same. These loose requirements on callers are possible because the function accepts a broad range of data and uses runtime checks before accessing fields or adding values.

**Challenge: Typing Dynamic Control-Flow**   Reasoning about such code is a challenge for static type systems, and calls for a notion of *type narrowing*. A type narrowing system introduces at least three high-level features: it adds a degree of flow sensitivity, as conditional tests may refine the type environments in their positive and negative branches; it adds a notion of paths into a data structure, such as the field rainfall in a JSON object; and it extends the language of types to express unions and subtractions without overwhelming complexity. A typechecker must have a logic that satisfies all of the above requirements to typecheck the avg_rainfall function.

With so many constraints at play, it is no surprise that it took decades of research, from Typed Lisp [8] and StrongTalk [4] to soft [77] and gradual [39, 43, 61, 69] typing, before type narrowing became a practicable reality in gradual type systems such as TypeScript [2], Flow [15], and Typed Racket [71]. And, naturally, these type systems disagree on several aspects of type narrowing—some incidental, due to variations among the languages they target, and some fundamental. For instance, Typed Racket has a nonstandard typing relation that assigns one type and two logical implications to every expression [68]. One benefit of this extra machinery is that Typed Racket can understand the logic in the following program, whereas TypeScript and Flow do not see that y must have type String in the return clause:

```
1  define nested_condition(x: Top, y: Top):
2    if (if x is Number: y is String else: false)
3      return x + String.length(y)  // expected: no type error, x is a number and y is a string
```

A natural question is whether the nonstandard judgment form of Typed Racket is worth the effort, as the new forms of composition and abstraction that it enables come at the price of a more complex typechecker, with negative implications for users and for type system maintainers.

As researchers, we cannot provide a definitive answer. Language communities must weigh the benefits in light of their own priorities and resources. We can, however, facilitate informed decisions by contributing a *language design benchmark* that clarifies the design space of type narrowing systems. That is the goal of this paper.

**Contributions**   We present a benchmark for type narrowing to showcase core behaviors that *should* and *should not* typecheck as a guide for language designers and a reference for researchers. It is termed a benchmark because its purpose extends beyond testing correctness in an absolute sense; rather, the goal is to characterize the abilities of narrowing implementations and highlight their fundamental tradeoffs. The benchmark is grounded in both the literature on type narrowing and the documentation of widely-used languages. We call the benchmark *If-T* to reflect its focus on conditional type tests (*If* for if/else, *T* for types).

If-T (current version [31]) has the following characteristics:

- It highlights 13 features of type narrowing systems;
- Each feature is illustrated by two small programs, one that should typecheck and one that should not;
- It contains implementations of the programs for five typecheckers: TypeScript, Flow, mypy, Pyright, and Typed Racket; and
- Finally, beyond assessing core functionality, If-T presents challenge problems and discusses how orthogonal type system dimensions (such as subtyping) should cooperate with type narrowing.

If-T is available online. We welcome discussions, issue reports, and pull requests:

https://github.com/utahplt/ifT-benchmark

Adding an implementation for a new typechecker is as simple as creating a folder, translating the benchmark programs, and specifying execution instructions in a common format (Appendix A). Finally, we provide a template datasheet (inspired by [26, 41]) for benchmark implementors to present their work in a uniform way, facilitating cross-language comparisons. All scripts for running the benchmarks and collecting the results are in the repository and the accepted artifact [31].

This paper is organized as follows. Section 2 discusses the challenges and tradeoffs of type narrowing. Section 3 presents our method for creating the benchmark. Section 4 presents the core If-T benchmark in detail. Section 5 presents the results of the benchmark on five typecheckers. Section 6 describes example programs that use the core functionality of type narrowing in compelling ways. Section 7 presents the datasheet template for summarizing implementations. Section 8 presents related work. Section 9 concludes with a brief discussion.

## 2　Background: Type Narrowing

Type narrowing is a flow-sensitive technique for typing code that handles dynamically-typed data. To illustrate, consider an if statement that inspects a value of type JSON:

```
1  // val :: JSON = Object | Array | String | Number | true | false | null
2  if val is Number and val ≤ 999:
3     total += val
```

This code assumes an *untagged union type* for JSON values that directly matches the JSON specification [19]: a JSON value can be an object, or an array, or one of several basic kinds of values. Unlike a *tagged union type*, these possibilities are not distinguished by type constructors. The code tests whether val is a number and then proceeds to use it in a comparison (≤) and an addition (+=). A typechecker thus needs to understand that type tests can narrow the types of JSON values.

　　Conventional type systems do not support type narrowing or untagged unions. In OCaml, for example, the way to represent JSON data is through a tagged datatype that labels each possibility with a separate constructor [56]. Instead of type tests, pattern matching is the standard way to check and untag values:

```
1  type json = `Object of (...) | `Int of int | ...  (* https://dev.realworldocaml.org/json.html *)
2  match val  (* val :: json *)
3  | `Int n when n <= 999 -> total := !total + n
4  | _ -> ()
```

Adding tags that classify data is straightforward work, and has organizational benefits (e.g., if the datatype changes, match errors will guide client changes), but programs written in dynamic languages rarely follow this idiom. Possible reasons include greater flexibility and code reuse, lower up-front design costs, or ignorance of tagging as a pattern. Whatever the reason, supporting this style of code is critical to allow gradual adoption of types without forcing sweeping changes upon a codebase.

　　Over the years, the need for type narrowing has been identified in several contexts [7, 32, 57]. Early solutions include conditional soft types [1, 76], multi-methods [6, 18], and CFA [33, 60]. The search for a precise and efficient solution has proven to be a longstanding challenge; below we review some highlights of the journey so far.

**Set-Theoretic Types**　　Semantic set-theoretic types enable a direct form of type narrowing. These types include syntax for exactly the necessary ingredients: operators for union, intersection, negation, and subtraction. These operators are supported by a semantics that represents types as sets of possible inhabitants:

$$\tau \quad ::= \quad \dots \mid \tau \cup \tau \mid \tau \cap \tau \mid \neg \tau \mid \tau \setminus \tau$$

A type such as JSON above is directly expressible as the union of several alternatives, and the types for two branches of an if statement can intersect or subtract as needed, depending on the condition:

```
1  // val :: T
2  if val is Number:
3    ... // val :: T ∩ Number
4  else:
5    ... // val :: T \ Number
```

User-defined functions can refine types as well. Generalizing from the test above (val is Number), type-refining predicates produce a boolean result and should have the side effect of modifying the subsequent type environment. This can be achieved by splitting a function type into two cases, a positive case that returns True and a negative case that returns False:

```
1  (λ x. x is Number) :: ∀ T. [Number -> True] ∩ [T \ Number -> False]
```

With this design, a typechecker can work backwards from the "then" branch of a conditional to refine its environment using the domain of the predicate function in the True case. Several typecheckers use this backwards-reasoning approach to handle type narrowing quite precisely [9, 10, 11, 12], though the use of predicates that carve out infinite subsets (such as is_even) tend to cause performance issues.

As a crowing example, a recent set-theoretic typechecker [11] based on CDuce [13] is able to *infer* types without any user-provided annotations. The following example code, from [11], describes a flatten function. Flatten reduces an arbitrarily-nested list to a flat sequence, e.g., flatten([[a], b, [[c,d]]]) = [a,b,c,d]. Flatten also wraps any non-list input in a flat list. The inferred type, in OCaml comments (between (* *) markers), covers both possibilities:

```
1  (* type Top = [anything, Top covers the entire universe of values]
2   * type Nested = List(Nested) ∪ (A \ List(Top))
3   *
4   * flatten :: Nested -> List(A \ List(Top))
5   *        ∩ (B \ List(Top)) -> List(B \ List(Top)) *)
6  let rec flatten t = match t
7    | [] -> []
8    | hd::tl -> concat (flatten hd) (flatten tl)
9    | _ -> [t]
```

Unfortunately, the typechecker takes over 300 seconds to infer a type for flatten. Scaling set-theoretic types to quickly checkr multi-million line codebases, as typecheckers for Python [40] and JavaScript [15] routinely do, presents a significant research challenge. Even an overloaded type for addition that matches the semantics of Racket can lead to slowdowns on the order of several minutes to typecheck simple uses [36, Appendix C]. Ongoing work to type Elixir [9] may bridge the gap to practice. Luau is one mainstream language that incorporates set-theoretic types, as a fallback in specific cases [35]. Other typecheckers rely on faster but less-precise syntactic techniques, such as occurrence typing.

**Occurrence Typing**    Typed Racket uses a syntactic technique called *occurrence typing* to narrow types [68, 70]. Occurrence typing has proven to be quite expressive, efficient, and extensible (e.g., [14, 37]), but it requires deep changes to the typechecking process. Whereas standard type inference uses a term ($e$) and an environment ($\Gamma$) to synthesize a type, occurrence typing synthesis four outputs: a type ($\tau$), positive and negative propositions ($\phi_+, \phi_-$), and an objective path ($o$, called an *object* in [68]):

$$\frac{\vdots}{\Gamma \vdash e : \tau ; \phi_+ \mid \phi_- ; o}$$

The two propositions describe how to modify the environment for the true and false branches of a conditional when this expression is used as the test. Every expression needs these propositions because in a dynamic language every kind of expression is valid to use in a conditional test. The objective is an optional path into a data structure (e.g., "first element of a pair") that tells what part of a data structure the expression inspects. Our running example val is Number would use an empty path. The variant vals[0] is Number would have a path into the first element of the array vals.

With this reformulated judgment form, only a few additional changes are needed. Types ($\tau$) must include untagged unions, environments ($\Gamma$) must track propositions in addition to types, and the typechecker must have metafunctions that compute syntactic approximations of type subtraction and intersection. These changes are modest relative to what true set-theoretic types require, and yet, as our benchmark shows, they enable fairly precise typing.

**Ad-Hoc Narrowing**    Both occurrence typing and set-theoretic types call for a steep investment in typechecker infrastructure. A natural question is whether the investment really pays off, especially since common-case narrowings can be supported with rudimentary syntactic checks. To support our running test, **if** val is Number: …, a typechecker merely needs to spot the phrase "is Number" and refine the type of the variable val. This sort of narrowing does not require negative propositions, objectives, or the complicated arrow types that appear in Typed Racket. If the test is applied to an expression that falls outside the common case, say f(val) is Number, then ad-hoc narrowing cannot help.

TypeScript, Flow, Pyright, mypy, and several other typecheckers implement ad-hoc narrowing rather than a compositional solution. As we will see, these systems have a variety of limitations. Some allow user-defined predicates, but do not typecheck them. Some are robust against simple program transformations; others are not. The If-T Benchmark is designed to lay these differences on the table so that language users can decide what features really matter for everyday programming.

## 3   Benchmark Design

If-T draws on two main sources for type narrowing examples, encompassing dozens of research articles and many pages of language documentation:

- *Research literature*. Section 2 cites significant sources. In addition, Tobin–Hochstadt and Felleisen's [68] sequence of motivating examples for occurrence typing provides a strong starting point, and both Fagan [20] and Greenberg [28] identify flatten as a challenge problem.
- *Typechecker implementations*, including their documentation [45, 46, 51, 52, 72, 74, 78], and online discussions, e.g., [49, 75]. We focused on TypeScript [73], Flow [44], mypy [64], and Pyright [50] because they have wide adoption.

First and foremost, these sources provide examples of what type narrowing ought to achieve, according to a variety of language communities. Second, these sources provide insights regarding type system features that enable narrowing in an implementation. We used these inputs to identify a set of core dimensions for type narrowing, detailed in Section 3.1. Along the way, we encountered topics that impinge on the correctness of type narrowing, but not on its core functionality. Subtyping, for example, has an impact on *what* information type narrowings can learn, but not on *how* that information gets stored. Section 3.2 lists these closely-related, but orthogonal, dimensions.

## 3.1 Core Dimensions

At a high level, there are four technical dimensions of type narrowing. The first, basic dimension is: (1) refine types based on conditional tests. Upon this foundation, narrowing systems should: (2) focus refinements on elements within a data structure, (3) soundly propagate information through program control flow, and (4) allow and typecheck user-defined predicates. We survey these dimensions in turn.

**Basic Narrowing**   At the most basic level, type narrowing systems must use type tests to refine types in conditional branches. Refinements must work in two directions, positive and negative, for the "then" and "else" branches of an if.

Equivalently, logical connectives such as negation (not), conjunction (and) and disjunction (or) should refine types in a similar way. For example, the negation of a type test should work analogously to the original type test, and a conjunction with type tests as its conjuncts should refine types in following conjuncts:

```
define maybe_add(n : Number | None):
    if not (n is None):
        return n + 1
```

```
define positive_number(n: Number | None):
    // comparison is type-safe
    return (n is Number) and (n > 0)
```

The language of type tests must match variables against basic types such as Number, String, and Boolean to meet this basic level of support.

**Compound Structures**   Extending the language of type tests to describe compound data structures is a second dimension of narrowing. Elements of fixed-size and arbitrary-size structures, such as tuples and lists, should be targets for refinements:

```
define fst_add(pair : Tuple(Top, Top)):          define list_add(xs : List(Top)):
    if pair[0] is Number:                            if xs is List(Number):
        return pair[0] + 1                               return sum(xs)
```

Same goes for fields in objects or records, as the rainfall example from Section 1 illustrates (if day["rainfall"] is Number: …).

**Advanced Control Flow**    Generalizing from the basic example of if/else branches, any path through a program should, in principle, support type narrowing. Multi-way conditionals, nested ifs, and loops that contain type tests might refine the environment. Assert statements, which halt the program when a test fails, are another example. This dimension of narrowing calls for careful management of the environment during typechecking. Along the same lines, storing the value of an expression in a variable should not inhibit narrowing:

```
// xs :: List(Top)              assert x is Number          tmp = x is String
for x in xs:                    x + 1                        if tmp:
    if not (x is String):                                       String.length(x)
        return False
// xs :: List(String)
```

**Custom Predicates**    When program expressions can work together to form predicates that go beyond the basic "var is Type" questions, users need the ability to use standard, functional abstractions to craft predicates that can be reused and unit-tested. User-defined predicates should have the same power to narrow types as inlined code. At a minimum, this calls for a sort of type annotation that describes a type-narrowing predicate. Since annotations can fall out of sync with code, it is critical that the type system validates predicates against their annotation. For example, the annotation below describes a symmetric predicate—the return type "x is Number" (from TypeScript) means that this function returns True only when x is a number and False otherwise. Its implementation, however, is not symmetric. When the function returns False, it is not safe to conclude that the input x is not a number:

```
1  define is_even(x : Top) -> x is Number:
2      return (x is Number) and (x mod 2 == 0)
3
4  is_even(4)  // True
5  is_even("A") // False
6  is_even(3)  // False
```

This is_even function should fail to typecheck because it does not decide whether its input is a number. A better return type is "implies x is Number" (which Flow supports but TypeScript does not), meaning that x is a number if the function returns true.

### 3.2 Orthogonal Dimensions

Type narrowing systems cannot refine types when it is unsound to do so. In a full-featured programming language, several factors can complicate the soundness picture. We describe common issues below.

**Subtyping**    All type tests in this paper, and in the If-T benchmark, use the keyword "is" to describe a type equality test. In a language with subtyping, however, type equality can give surprising results. For example, the following Python code is unsound despite seeming correct:

```python
def f(x: str | int | bool) -> int:
    if (not type(x) is str) and (not type(x) is bool):
        return x + 1
    else:
        return 0
```

This is because the type str is extensible by subclasses, so calling f with an instance of a subclass of str could cause a runtime error. Subtyping tests, such as isinstance in Python and instanceof in JavaScript, are more idiomatic. This concern falls outside the main focus of If-T. The benchmark uses subtyping only to narrow atomic types from unions of atomic types.

**Mutation**    Mutation can change the shape of values and thereby invalidate observations made by earlier type narrowings. An important aspect of handling control flow in a typechecker is thus to identify write operations and narrow (or rather, un-narrow) types accordingly. Discovering writes precisely is an orthogonal research direction.

For example, in Python, a field access obj.f can run user-defined code if the object has a property method named f. Thus, the second field access below is potentially unsafe because the evaluation of self.parent.wins may replace self.parent with None:

```python
if self.parent is not None:
    total += self.parent.wins
    total += self.parent.losses  # may be unsound!
```

Interestingly, neither mypy nor Pyright detect this issue. Instagram's Static Python language conservatively flags an error [29, 42]. Detecting error cases precisely is difficult. Property methods might be inherited from legacy Python objects that are not analyzed by the typechecker.

**Concurrency**    When combined with mutation and aliasing, language support for concurrency makes it even more difficult to determine what code can safely rely on the results of a type test. Typed Racket never allows narrowing on mutable data because the object might be shared across threads.

■ **Table 1** If-T Benchmark Items

| | | |
|---|---|---|
| | Basic Narrowing: | |
| 1. | positive | Refine when condition is true |
| 2. | negative | Refine when condition is false |
| 3. | connectives | Handle logic connectives: not, or, and |
| 4. | nesting_body | Conditionals nested within branches |
| | Compound Structures: | |
| 5. | struct_fields | Refine fields of immutable structure |
| 6. | tuple_elements | Refine tuple elements |
| 7. | tuple_length | Refine based on tuple size |
| | Advanced Control Flow: | |
| 8. | alias | Track logical implication through variables |
| 9. | nesting_condition | Conditionals nested within conditions |
| 10. | merge_with_union | Correctly merge control-flow branches |
| | Custom Predicates: | |
| 11. | predicate_2way | Custom predicates that narrow positively and negatively |
| 12. | predicate_1way | Custom predicates that narrow only positively |
| 13. | predicate_checked | Typecheck the body of custom predicates |

## 4 The Benchmark

https://github.com/utahplt/ifT-benchmark/blob/main/README.md#the-benchmark

The If-T benchmark consists of 13 items, each focusing on a different feature of type narrowing systems. These items are divided into four groups to match our core conceptual dimensions (Section 3.1). Each item comes with a short description and is illustrated with "Success" and "Failure" code examples. Success code should typecheck. Failure code should raise a specific type error.

Table 1 provides brief overview of If-T by listing the name and description for each benchmark item. As the table shows, items are evenly divided across the four categories. The subsections below first discuss the pseudocode syntax used in the benchmark, then provide full example programs.

### 4.1 Pseudocode Overview

If-T is written in a pseudocode. To implement the benchmark, this pseudocode must be adapted to a real target language (JavaScript, Ruby, etc.). We offer the following points to guide translation efforts:

- Functions are defined with type annotations for the parameters and the return type, e.g., define f(x: T) -> U:.
  - A return type of the form "x is T" expresses a symmetric predicate. Well-typed implementations must return a Boolean with the value True when x has the type T and the value False when x does not have the type T.

– A return type "implies x is T" expresses an asymmetric predicate. Well-typed implementations must return a Boolean with the value True only if x has the type T. There are no other conditions.

- Control flow is expressed using if/else blocks. These can appear anywhere that an expression is expected, including in the test of another if/else expression.
  – Languages that categorize if/else as a statement form may have a ternary conditional operator.
- Type tests use the form x is T.
- let binds an immutable variable. var binds a mutable variable.
- The pseudocode types Number, String, and Boolean are disjoint and final:
  – Adding a Number to a String is an error, unlike in JavaScript.
  – Subclassing String is not allowed, unlike in Python with str.
- The type Top is the superclass of every other type.
- There are no explicit type casts in the pseudocode.
- The struct keyword defines a tuple with named elements. Elements are accessed using dot notation (x.f).

## 4.2 Basic Narrowing

There are four parts in the Basic Narrowing section: positive, negative, connectives, and nesting_body.

### 4.2.1 Positive

If a type test succeeds on a variable, the type of the variable is refined to a more-specific type based on the predicate. The example programs apply the test x is String. The first program treats the result as a string, and should typecheck. The second treats the result as a number, and should fail to typecheck.

**positive: Success**

```
define f(x: Top) -> Top:
   if x is String:
      return String.length(x)
   else:
      return x
```

**positive: Failure**

```
define f(x: Top) -> Top:
   if x is String:
      return x + 1
   else:
      return x
```

### 4.2.2 Negative

If a type test fails on a variable, the type of the variable is refined to a more-specific type based on the negation of the predicate in the appropriate control path. These programs also use the test x is String:

---

**negative: Success**

```
define f(x: String | Number) -> Number:
    if x is String:
        return String.length(x)
    else:
        return x + 1
```

---

**negative: Failure**

```
define f(x: String | Number | Boolean) -> Number:
    if x is String:
        return String.length(x)
    else:
        return x + 1
```

---

### 4.2.3 Connectives

A predicate built from several type tests and logical connectives should result in narrowing that matches the connectives. For a conjunction of tests applied to the same variable, the result type should be the intersection of each tested type. For a disjunction of tests, the result type should be the union of each type. For a negation, the result type should use the complement of the tested type.

---

**connectives: Success**

```
define f(x: String | Number) -> Number:
    if not (x is Number):
        return String.length(x)
    else:
        return 0

define g(x: Top) -> Number:
    if x is String or x is Number:
        return f(x)
    else:
        return 0

define h(x: String | Number | Boolean) -> Number:
    if not (x is Boolean) and not (x is Number):
        return String.length(x)
    else:
        return 0
```

---

**connectives: Failure**

```
define f(x: String | Number) -> Number:
    if not (x is Number):
        return x + 1
    else:
        return 0

define g(x: Top) -> Number:
    if x is String or x is Number:
        return x + 1
    else:
        return 0

define h(x: String | Number | Boolean) -> Number:
    if not (x is Boolean) and not (x is Number):
        return x + 1
    else:
        return 0
```

#### 4.2.4 Nesting Body

When a conditional statement is nested inside the body of another conditional statement, the type of the variable is refined to the intersection of the types refined by each conditional statement.

**nesting_body: Success**

```
define f(x: String | Number | Boolean) -> Number:
    if not (x is String):
        if not (x is Boolean):
            return x + 1
        else:
            return 0
    else:
        return 0
```

**nesting_body: Failure**

```
define f(x: String | Number | Boolean) -> Number:
    if x is String | Number:
        if x is Number | Boolean:
            return String.length(x)
        else:
            return 0
    else:
        return 0
```

## 4.3 Narrowing with Compound Structures

There are three parts in the Compound Structures section: struct_fields, tuple_elements, and tuple_length.

### 4.3.1 Struct Fields
When a predicate is applied to a field of an immutable data structure, the type of that property is narrowed. Other field types (if any) remain the same.

**struct_fields: Success**

```
struct Apple:
    a: Top

define f(x: Apple) -> Number:
    if x.a is Number:
        return x.a
    else:
        return 0
```

**struct_fields: Failure**

```
struct Apple:
    a: Top

define f(x: Apple) -> Number:
    if x.a is String:
        return x.a
    else:
        return 0
```

### 4.3.2 Tuple Elements
When a predicate is applied to an element of a tuple, refine the type of that element. Similar narrowings should work for other covariant positions such as the elements of a list or the return value of a function.

**tuple_elements: Success**

```
define f(x: Tuple(Top, Top)) -> Number:
    if x[0] is Number:
        return x[0]
    else:
        return 0
```

**tuple_elements: Failure**

```
define f(x: Tuple(Top, Top)) -> Number:
    if x[0] is Number:
        return x[0] + x[1]
    else:
        return 0
```

### 4.3.3 Tuple Length

Because tuple types list the type of each element, they also describe the length of the overall tuple. Thus, if a variable may point to different-sized tuples, type narrowing based on the length of its value should refine its type. Similar narrowings should work for other observable properties that data-structure types describe.

**tuple_length: Success**

```
define f(x: Tuple(Number, Number) | Tuple(String, String, String)) -> Number:
    if Tuple.length(x) is 2:
        return x[0] + x[1]
    else:
        return String.length(x[0])
```

**tuple_length: Failure**

```
define f(x: Tuple(Number, Number) | Tuple(String, String, String)) -> Number:
    if Tuple.length(x) is 2:
        return x[0] + x[1]
    else:
        return x[0] + x[1]
```

### 4.4 Advanced Control Flow

There are three parts in the Advanced Control Flow section: alias, nesting_condition, and merge_with_union.

### 4.4.1 Alias

When the result of a predicate test is bound to an immutable variable, that variable can also be used as a type guard. When the result of a predicate test is bound to a mutable variable, that variable can be used as a type guard only if it is not updated.

**alias: Success**

```
define f(x: Top) -> Top:
    let y = x is String
    if y:
        return String.length(x)
    else:
        return x
```

---

**alias: Failure**

```
define f(x: Top) -> Top:
    let y = x is String
    if y:
        return x + 1
    else:
        return x

define g(x: Top) -> Top:
    var y = x is String // y is mutable
    y = true
    if y:
        return String.length(x)
    else:
        return x
```

---

### 4.4.2 Nesting Condition

When a conditional statement is nested inside the condition of another conditional statement, the type of the variable is refined in the same manner as if the tests were joined by logical connectives. In the following programs, the first nested if statement effect a conjunction of two type tests, and the second one first checks if x is a number and then checks if y is a string regardless of the result of the first check.

---

**nesting_condition: Success**

```
define f(x: Top, y: Top) -> Number:
    if (if x is Number: y is String else: false):
        return x + String.length(y)
    else:
        return 0
```

---

**nesting_condition: Failure**

```
define f(x: Top, y: Top) -> Number:
    if (if x is Number: y is String else: y is String):
        return x + String.length(y)
    else:
        return 0
```

---

### 4.4.3 Merge with Union

If a variable has different types in two branches of a conditional and control-flow after these branches merges to a common point, then the variable at that point should take the precise union of the two types. The variable's type should not conservatively change to the common supertype (Top) at the merge point.

**merge_with_union: Success**

```
define f(x: Top) -> String | Number:
    if x is String:
        String.append(x, "hello")
    else if x is Number:
        x = x + 1
    else:
        return 0
    return x
```

**merge_with_union: Failure**

```
define f(x: Top) -> String | Number:
    if x is String:
        String.append(x, "hello")
    else if x is Number:
        x = x + 1
    else:
        return 0
    return x + 1
```

## 4.5 Narrowing with Custom Predicates

There are three parts in the Custom Predicates section: predicate_2way, predicate_1way, and predicate_checked. The first two test caller-side uses of predicates. The third tests the validity of predicate definitons.

### 4.5.1 Predicate 2-Way

When a symmetric (or, 2-way) predicate is true, the type of the variable is refined to a more-specific type with the information that the predicate holds. When a symmetric predicate is false, the type of the variable is refined in the opposite way.

**predicate_2way: Success**

```
define f(x: String | Number) -> x is String:
    return x is String

define g(x: String | Number) -> Number:
    if f(x):
        return String.length(x)
    else:
        return x
```

---

**predicate_2way: Failure**

```
define f(x: String | Number) -> x is Number:
    return x is Number

define g(x: String | Number) -> Number:
    if f(x):
        return x
    else:
        return x + 1
```

---

### 4.5.2 Predicate 1-Way

When a 1-way, positive predicate is true, the type of the variable is refined to a more-specific type with the information that the predicate holds. When a positive predicate is false, the type of the variable is not refined. Positive predicates typically check an underapproximation of a static type. In the example code, the predicate checks for a nonnegative number but narrows its input type to merely `Number`.

---

**predicate_1way: Success**

```
define f(x: String | Number) -> implies x is Number:
    return x is Number and x > 0

define g(x: String | Number) -> Number:
    if f(x):
        return x + 1
    else:
        return 0
```

---

**predicate_1way: Failure**

```
define f(x: String | Number) -> implies x is Number:
    return x is Number and x > 0

define g(x: String | Number) -> Number:
    if f(x):
        return x + 1
    else:
        return String.length(x)
```

---

### 4.5.3 Predicate Checked

A typechecker should confirm that the body of a custom predicate matches its type annotation. It should not blindly trust the annotation.

---

**predicate_checked: Success**

```
define f(x: String | Number | Boolean) -> x is String:
    return x is String

define g(x: String | Number | Boolean) -> x is Number | Boolean:
    return not f(x)
```

---

**predicate_checked: Failure**

```
define f(x: String | Number | Boolean) -> x is String:
    return x is String or x is Number // may return true when predicate is false

define g(x: String | Number | Boolean) -> x is Number | Boolean:
    return x is Number // may return false when predicate is true
```

---

## 5　Benchmark Results

Table 2 shows the results of the If-T benchmark on five full-featured typecheckers: Typed Racket, TypeScript, Flow, Mypy, and Pyright. (Research languages for set-theoretic types are not included at this time, but we would welcome future contributions.) A circle (●) indicates that the system passes the two examples of the benchmark item by validating the Success example and catching the Failure example.

■ **Table 2**　Benchmark Results: ● = passed, ✕ = failed imprecisely, ⬚✕ = failed unsoundly

| | | Typed Racket | TypeScript | Flow | Mypy | Pyright |
|---|---|:---:|:---:|:---:|:---:|:---:|
| | Basic Narrowing: | | | | | |
| 1. | positive | ● | ● | ● | ● | ● |
| 2. | negative | ● | ● | ● | ● | ● |
| 3. | connectives | ● | ● | ● | ● | ● |
| 4. | nesting_body | ● | ● | ● | ● | ● |
| | Compound Structures: | | | | | |
| 5. | struct_fields | ● | ● | ● | ● | ● |
| 6. | tuple_elements | ● | ● | ● | ● | ● |
| 7. | tuple_length | ✕ | ● | ● | ● | ● |
| | Advanced Control Flow: | | | | | |
| 8. | alias | ● | ● | ✕ | ✕ | ● |
| 9. | nesting_condition | ● | ✕ | ✕ | ✕ | ✕ |
| 10. | merge_with_union | ● | ● | ● | ✕ | ● |
| | Custom Predicates: | | | | | |
| 11. | predicate_2way | ● | ● | ● | ● | ● |
| 12. | predicate_1way | ● | ✕ | ● | ● | ● |
| 13. | predicate_checked | ● | ⬚✕ | ● | ⬚✕ | ⬚✕ |

**If-T: A Benchmark for Type Narrowing**

A cross indicates a failure on either the Success or Failure example. There are two styles of cross: a cross without a red background (✗) indicates imprecise typing, and a cross with a red background (❌) indicates unsound typing. No language has full support for the benchmark, and the differences across languages lead to several interesting observations.

**Basic Narrowing**    Every typechecker has full support for basic narrowing features (positive, negative, connectives, nesting_body). These core functionalities are well-established and agreed on in mainstream type narrowing systems.

**Compound Structures**    Every typechecker is able to narrow data-structure elements. With the exception of Typed Racket, every typechecker is able to narrow tuples based on their size (tuple_length). Adding support for direct size tests is an area for improvement in Typed Racket; currently, code must use primitive elimination forms (car, cadr), as illustrated by the following well-typed program:

```
1  (define (tuple-length-success-f [x : (U (List Number Number) (List String String String))])
2    (if (null? (cdr (cdr x))) ;; check: tuple length = 2
3        (+ (car x) (cadr x))
4        (string-length (car x))))
```

**Advanced Control Flow**    There are major differences among typecheckers in the advanced control flow category:

- Flow and mypy ignore type tests that are bound to variables (alias). Tests must appear directly in a conditional, which may lead to duplicated code.
- Only Typed Racket supports tests within a conditional test (nesting_condition). This difference is partly due to Racket's emphasis on compositional syntax: in Racket, if is an expression form, whereas in JavaScript and Python if is a statement. To add support, the other languages would need an expression-typing judgment that tracks positive and negative propositions like occurrence typing (Section 2).
- Mypy is overly conservative when merging branches (merge_with_union). Instead of merging String and Number to a union, it jumps to Top. This problem is well-known in the mypy community. It has a dedicated issue tag in the mypy repository that is currently applied to 49 open issues [53].

**Custom Predicates**    Support for user-defined predicates is another contentious topic. Every typechecker allows the declaration of symmetric predicates (predicate_2way), but differences abound on the finer details:

- TypeScript does not allow asymmetric predicates, such as the is_even example above (Section 3.1). Every TypeScript predicate must refine types positively when it returns true and refine types negatively when it returns false; a predicate can never "not know" about an input.
- TypeScript, mypy, and Pyright do not check the soundness of user-defined predicates. Nonsensical predicates that lead to unsafe type coercions are permitted, such as the following Python function that casts any input to have a function type. Mypy reports no type errors:

```python
def f(x) -> TypeIs[Callable[[int], int]]:
    return True
```

```python
a = 42
if f(a):
    a(1) # runtime error
```

## 5.1 Additional Observations

Although Flow can typecheck user-defined predicates, it supports only a limited syntax. Predicates can contain just a single expression. The following predicate, with a simple if statement, is not allowed:

```
1  function h(x: string | number): x is string {
2    if (typeof(x) == "string") { return true } else { return false }
3  } // Error: consider replacing the body of this predicate function with a
4    //  single conditional expression.
```

Typed Racket has extensive support for user-defined predicates. It allows asymmetric predicates that refine only in the positive case or only in the negative case. Its predicates can refine any subset of their arguments and any objective path into those arguments. Mypy and Pyright, by contrast, can refine only the first argument [72, 78]. These possibilities are enabled through a domain-specific language embedded in function return types [55]. Does this expressiveness lead to any practical returns? We would expect so, but none of the libraries included with Typed Racket use its keywords for negative asymmetry or objective paths (#:-, #:object). We have found only one project that uses negative predicates (for the Pie prover [23]) and it simply says that when a predicate for variable names returns false, the input must be a keyword [16].

Finally, we must acknowledge that Typed Racket's predicates are not perfectly sound. Typed code is allowed to import a dynamically-typed function $f$ and assign it a positive predicate type, via the #:opaque keyword. If $f$ uses state, its use can put contradictory information in the type environment [38]. Preventing abuses would require heavy runtime machinery, which brings us back to the safety and performance tradeoffs inherent to gradual typing. Perfect soundness is unrealistic, and yet typecheckers can achieve much more than the light-touch approach of TypeScript. What is the right balance to strive for?

## 6  Example Programs

https://github.com/utahplt/ifT-benchmark/blob/main/EXAMPLES.md

In addition to the core benchmark, If-T includes a set of example programs. The purpose of the examples is to show how the features of type narrowing come together to support useful and practical programs. Each example comes with a variant that has a relevant type error, to catch systems that unsoundly trust user annotations.

▪ **Table 3**  Example Programs and the core benchmark items they depend on

```
filter     ← positive + predicate_2way (or _1way) + tuple_elements (for lists)
flatten    ← positive + negative
TreeNode   ← positive + negative + predicate_checked + nesting_body
Rainfall   ← positive + object_properties + nesting_body
```

The covered features for each example are listed in Table 3, which also aligns with our interpretation of the benchmark result: basic narrowing features like positive and negative are used in most examples, while more advanced features play important roles for implementing real-world program logic.

We introduced two of the examples earlier in this paper to motive narrowing: Rainfall (Section 1) and flatten (Section 2). These rely on nested conditional branches, compound structures, and aliasing. There are two other examples: TreeNode, a recursive user-defined predicate; and filter, a higher-order function that takes a predicate as input to refine a data structure:

```
1  define filter(predicate: (x: T) -> x is S, list: List(T)) -> List(S)
2    let result = []
3    for element in list:
4      if predicate(element):
5        result = cons(element, result)
6    return result
```

As a historical aside, the creators of Typed Racket used Filter as an early milestone [70]. When they could type it, they knew their design was on the right track (personal communication with Felleisen). TypeScript added support for anonymous Filter predicates in TypeScript 5.5 [58].

## 7  Datasheet Template

https://github.com/utahplt/ifT-benchmark/blob/main/DATASHEET.md

The final component of If-T is a datasheet that summarizes implementation efforts. It asks a series of fourteen open-ended questions, ranging from *What is the dynamic type in your language?* to *Are any benchmarks inexpressible? Why?*. The authors of an If-T implementation are to answer the questions using prose and links to existing resources (such as language documentation). The purpose of this exercise is to give readers of the benchmarks a high-level and relatively uniform overview of how each implementation approaches type narrowing. With this basic understanding in hand, readers should be better equipped to understand the code and to separate incidental choices (e.g., related to syntax) from the core aspects of type narrowing.

## 8 Related Work

If-T as a benchmarking effort is influenced by prior language-design benchmarks. B2T2, the Brown Benchmark for Table Types [41], sets criteria for type systems that support tabular data. It has six components: a baseline definition of tables, small example tables to encode, a full-featured API (largely inspired by the success of tabular programming in Pandas [65] and R [67]), example programs, error programs, and a datasheet to summarize implementations. If-T adapts similar components to the issue of type narrowing. Jakubovic et al. [34] propose technical dimensions of programming systems, such as notational structure, composability, and learnability, and broader clusters such as interaction and conceptual structure. These divisions into dimensions and clusters influenced our characterization of type narrowing (Table 1). Other benchmark suites, from the Gabriel benchmarks [24] to the Renaissance suite [54], provided further inspiration [17, 27, 30, 59, 66].

## 9 Conclusion

Type narrowing is an important aspect of reasoning about dynamically-typed code. Dynamic programs lack the rigid organization that static types provide, and thus gravitate toward flexible check-driven control flow and duck typing. Proposed designs for type narrowing span a range of possibilities, from precise set-theoretic types to the deliberate unsoundness of TypeScript. The If-T benchmark provides a first, rigorous basis for language designers to compare the expressiveness and soundness of type narrowing systems. It is a *design benchmark* in the sense that it aims to characterize type systems by their precision and soundness, but does not mandate that perfect results are necessary. The benchmark consists of a core benchmark (with 13 pairs of programs), practical examples, and a summarizing datasheet to facilitate comparisons. The results on five typecheckers show considerable variation in core behavior (Section 5), substantiating the need for a benchmark. There are many other typecheckers with some form of narrowing, including Hack [47], Luau [5, 35], Pyre [48], Sorbet [63], Static Python [42], and Typed Clojure [3]. Typed languages might benefit from narrowing as well, e.g., to increase the flexibility of polymorphic variants [25]. We hope this work inspires and facilitates transfer of type narrowing ideas across languages.

**Data-Availability Statement**
The If-T benchmark and scripts for reproducing the results are in our artifact [31].

**Acknowledgements**    We thank Sam Tobin-Hochstadt for insightful conversations, Fred Fu for conducting a related investigation and reporting a bug in our initial Typed Racket implementation, Eric Traut for explaining str subtyping in Python, Carl Meyer for discussing Static Python's approach to property methods, and Ashton Wiersdorf and Andrew McNutt for comments on earlier drafts.

## A Artifact Overview

The If-T benchmark is available online and archived on Zenodo [31]. The following is a brief guide on how to install, run, and contribute to the benchmark. More details can be found in the `SETUP.md` and `CONTRIBUTING.md` files in the repository.

### A.1 Installation

Driver code for the If-T benchmark is implemented in Racket. To run the benchmark, first install Racket and the typecheckers that support the If-T implementations that you wish to run. Some of the typecheckers require additional dependencies to be installed. These are declared in the package manager manifest files of each implementation directory.

### A.2 Running the Benchmark

After installing the dependencies, the benchmark can be run by executing the `main.rkt` script in the top level of the repository. This script will run the benchmark items on each of the implementations (currently, five typecheckers) and output the results in a table format. Several formats are supported for the output, including plain text, markdown, and LaTeX. For example, the following command runs the benchmark and outputs LaTeX:

```
$ racket main.rkt -f tex
```

```
Benchmark      & typedracket & typescript & flow & mypy & pyright \\
positive       & O           & O          & O    & O    & O       \\
negative       & O           & O          & O    & O    & O       \\
...
```

### A.3 Contributing

The If-T benchmark is open to contributions for adding new benchmark items, example programs, and implementations. To contribute to the benchmark, fork the repository and submit a pull request. If you wish to add a new typechecker to the benchmark, follow the structure of the existing implementation directories and their respective `_benchmark.rkt` files. Also, update the top level driver file `main.rkt` to include the new implementation.

**About the authors**

**Hanwen Guo** (hanwen.guo@utah.edu) is a PhD student at the University of Utah.
https://orcid.org/0009-0000-7118-2145

**Ben Greenman** (benjamin.l.greenman@gmail.com) is an assistant professor at the University of Utah.
https://orcid.org/0000-0001-7078-9287